\documentclass[preprint,aps]{revtex4}

\usepackage{graphicx}

\begin{document}

\newcommand{\beq}{\begin{equation}}
\newcommand{\eeq}{\end{equation}}
\newcommand{\ben}{\begin{eqnarray}}
\newcommand{\een}{\end{eqnarray}}
\newcommand{\bea}{\begin{array}}
\newcommand{\eea}{\end{array}}
\newcommand{\om}{(\omega )}
\newcommand{\bef}{\begin{figure}}
\newcommand{\eef}{\end{figure}}
\newcommand{\leg}[1]{\caption{\protect\rm{\protect\footnotesize{#1}}}} 

\newcommand{\ew}[1]{\langle{#1}\rangle}
\newcommand{\be}[1]{\mid\!{#1}\!\mid}
\newcommand{\mat}[3]{\langle{#1}\be{#2} {#3}\rangle}
\newcommand{\no}{\nonumber}
\newcommand{\etal}{{\em et~al }}
\newcommand{\geff}{g_{\mbox{\it{\scriptsize{eff}}}}} 
\newcommand{\la}{\lambda}
\newcommand{\lab}{\la\!\!\!\!{\makebox[1pt]{}^-}}
\newcommand{\Ds}{\Delta_s}
\newcommand{\dd}{\partial}
\newcommand{\rr}{{\bf r}}
\newcommand{\Out}{{\rm out}}
\newcommand{\In}{{\rm in}}
\newcommand{\dir}{\hat{\Omega}}
\newcommand{\da}[1]{{#1}^\dagger}
\newcommand{\cf}{{\it cf.\/}\ }
\newcommand{\ie}{{\it i.e.\/}\ }
\newcommand{\paragraphe}[1]{\vrule height1.5pt depth0ex width0pt\paragraph{#1}\vrule
height0pt depth1.5ex width0pt\hfil\break}
\newcommand{\premierparagraphe}[1]{\paragraph{#1}\vrule height0pt depth1.5ex width0pt\hfil\break}
\newdimen\mahauteur
\newdimen\malargeur
\newbox\maboite
\def\jeposecentre#1{\setbox\maboite=#1\mahauteur=\dp\maboite\malargeur=
\wd\maboite\divide\malargeur by 2\kern-\malargeur\raise\mahauteur#1}
\def\moitie{\kern.1em\raise.5ex\hbox{\the\scriptfont0 1}\kern-0.15em/
  \kern-0.1em\lower.5ex\hbox{\the\scriptfont0 2}}

\title{
Vacuum-field atom trapping \\ in a wide aperture spherical resonator.}

\author{Jean-Marc Daul and Philippe Grangier} 
\affiliation{Laboratoire Charles Fabry de l'Institut  d'Optique, \\
 F-91403 Orsay cedex - France}

\begin{abstract}

We consider the situation where a two-level atom is 
placed in the vicinity of the center 
of a spherical cavity with a large numerical aperture. The vacuum field at the 
center of the cavity is actually equivalent to the one obtained in a microcavity, 
and both the dissipative and the reactive parts of the atom's
spontaneous emission are significantly modified. 
Using an explicit calculation of 
the spatial dependence of the radiative relaxation rate and of 
the associated level shift, we
show that for a weakly excitating light field, the atom can be attracted to the 
center of the cavity by vacuum-induced light shifts. 

\end{abstract}

\maketitle


\section{Introduction}

Many theoretical and experimental work has been devoted during recent 
years to the so-called ``cavity QED" regime, where strong coupling is achieved 
between a few atoms and a field mode contained inside a microwave or optical 
cavity. In particular, it has been demonstrated that the spontaneous emission 
rate of an atom inside the cavity is different from its value in free 
space \cite{note,D,K,GRGH,GD,HHK,HCTF,JAHMMH,DIJM,ZLML,HF,MYM,JMD1}. This 
effect can be discussed from several different approaches, and here it will be 
basically attributed to a change of the spectral density of the modes of the 
vacuum electromagnetic field, which is due to the cavity resonating structure \cite{JMD1}. 
This approach is particularly convenient when the cavity does not have one 
single high-finesse mode, but rather many nearly degenerate modes, as it is the 
case in confocal or spherical cavities. More precisely, we will show that a ``wide 
aperture" concentric resonator using spherical mirrors with a large numerical 
aperture, can in principle change significantly the spontaneous emission rate
of an atom sitting close to the cavity center, 
even with moderate finesse. Similar result were already demonstrated, using 
either a spherical cavity \cite{HCTF,HF,JMD1} or ``hour-glass" modes in a confocal cavity 
\cite{MYM}. 

In such experiments, the atom has to sit within the active region volume, 
which is usually of very small size (of order $(10 \lambda)^3$ to $(100 
\lambda)^3$). In refs \cite{HF,MYM}, a possible solution was implemented by using a 
narrow atomic beam, and by reducing the cavity finesse in order to have an 
extended area in which a spherical wave is ``self-imaged" on itself. However, 
getting large effects will put more severe constraints both on the quality of the 
cavity and on the localisation of the atoms. A different way to implement the 
proposed scheme in a spherical cavity, that we would like to discuss in more 
detail here, is to use light-induced forces in order to attract the atom to the 
cavity center. A possible implementation could be to couple a light beam inside 
the cavity, and then to use the dipole force to hold the atoms in the right 
position, i.e., close to the cavity center. The effect of strong atom-cavity 
coupling on the dipole force has been studied theoretically \cite{MLG,SC}, and very 
interesting effects can be expected : since the atomic relaxation will be modified 
by the cavity, the balance between the trapping and heating effects of the dipole 
trap will be changed with respect to free space, which could result in an 
improvement of the trap itself.

A good understanding of these effects requires first to know the full space 
dependence of the cavity-induced damping and level shifts. In this letter, we 
will look at the situation where the atom lies close to the center of a spherical 
cavity with a large numerical aperture. We will show that large changes both in 
the atom damping rate and in its energy levels can be expected, even with a 
moderate cavity finesse, provided that the atom sits (relatively, but not 
extremely) close to the cavity center \cite{JMD1}. Moreover, we will show that for a weak 
excitating field, the atom can be 
trapped by vacuum-induced light shifts \cite{HBR,ESBS}, which create a force whose 
spatial dependence is related to the shape of the mode spectral density. 
In the following, we will assume that the cavity damping rate $\kappa$ is 
much larger than the free-space atom damping rate $\Gamma_{vac}$. In that 
case, the cavity still acts as a continuum with respect to the atomic relaxation. 
This will allow us to treat simply the atom-field coupling using frequency-
dependant coupling coefficients. For simplicity, we will also assume a weak 
excitation of the atom. This assumption is however not crucial, and possible 
extensions will be discussed at the end of the paper.

\section{Light-induced forces in the cavity QED regime}

Let us consider a neutral, slowly moving two-level atom in the dipole 
approximation. The Hamiltonian describing the atom-light coupling is :
\beq 
H = H_f + H_a - {\bf D . E}({\bf r})
\eeq
where $\bf D$ is the atom dipole moment and $\bf E$ is the electric field. 
The free field and atom hamiltonians are respectively 
$H_f = \sum _k \hbar \omega _k 
(n_k + 1/2)$ and $H_a = {\bf p}^2/(2m) + \hbar \omega _o S^{(z)}$. 
Here, $\bf r$ and $\bf p$ are the atom's position and momentum, $m$ is its mass,
and the dipole operator can be written ${\bf D} = 
{\bf d} (e^{i \omega _L t} S^{(+)} + e^{-i \omega _L t} S^{(-)})$, 
where $S^{(+)}$ and $S^{(-)}$ are the usual two-level rising 
and lowering operators in the frame rotating at the angular frequency $\omega _L$
of an externally applied laser 
(we have $S^{(z)} = (S^{(+)} S^{(-)} - S^{(-)} S^{(+)})/2$).
 The electric 
field operator is expanded as usual on a basis of orthogonal modes, but we do 
not specify that these modes should be plane waves. One has therefore :
\beq
{\bf E}({\bf r},t) = i \sum_k {\bf e}_k ({\bf r}) \; a_k(t) \;e^{-i\omega _k t} + h.c.
\eeq
where $k$ is a mode label which include polarization, and 
${\bf e}_k({\bf r})$ is the contribution of mode $k$ to the field at point 
${\bf r}$ close to the cavity center. 
It will be convenient to refer optical frequencies to 
a reference  $\omega _L$,  which can be the frequency 
of an externally applied laser as said above, and to define :
\beq
{\bf A}({\bf r},t) = i \sum_k {\bf e}_k ({\bf r}) \; a_k(t) \;
e^{-i (\omega _k - \omega _L) t}
\eeq
so that : 
\beq
{\bf E}({\bf r},t) = 
{\bf A}({\bf r},t) e^{-i \omega _L t} + {\bf A}^\dagger({\bf r},t) e^{i \omega _L t}.
\eeq

Moving to a frame rotating at the frequency $\omega _L$
and using the rotating wave approximation, the interaction part 
of the Hamiltonian can be written : 
\beq 
H_{int} =  - S^{(+)} {\bf d . A} - {\bf d . A^\dagger} S^{(-)} 
\eeq
From the expression of the hamiltonian one can simply get the force acting 
on the atom\cite{GA}~: 
\beq
{\bf F} ({\bf r},t) = \frac{d{\bf p}}{dt} = 
-{\bf \nabla} H =  {\bf \nabla}(S^{(+)}(t) \; {\bf d.A}({\bf r},t)) + h.c. 
\eeq
In this expression, the only space-dependant part is ${\bf e}_k ({\bf r})$,
so the force can also be written~:
\beq
{\bf F} ({\bf r},t) =  i \; S^{(+)}(t) \; \sum_k {\bf \nabla} [\, {\bf d}.{\bf e}_k ({\bf r}) ] 
\; a_k(t) \; e^{-i (\omega _k - \omega _L) t}   + h.c. 
\eeq

From the Hamiltonian, one gets the Heisenberg equations for the field 
operators $a_k$  \cite{CDG}, which can be integrated formally, yielding 
(${\bf d}.{\bf e}_k$ is taken real)~:
\beq
a_k (t)  = a_k (t_0) + \frac{{\bf d}.{\bf e}_k}{\hbar}
\int_0^{t-t_0} d\tau S^{(-)}(t-\tau) e^{i(\omega_k -\omega_L)(t-\tau)}
\eeq
Using this result in the definition of the fields operators, one sees that the 
field splits in two parts  ${\bf E} = {\bf E_o} + {\bf E_s}$, 
where ${\bf E_o}$ and ${\bf E_s}$ are the well-known 
``vacuum field" and ``source field" terms, which do not commute  \cite{CDG}.
Similarly, the expression of the force splits in two parts~: 
\beq
{\bf F}= {\bf F_o} + {\bf F_s}
\eeq
By choosing the normal ordering when separating the two non-commuting 
terms, the first part ${\bf F_o}$
gives the usual expression of the light-induced force \cite{GA}.  
The second term ${\bf F_s}$ is zero in the absence of a cavity, because the 
gradient of the source field is zero at the dipole place. However, as we will show now,
this is no longer true inside a cavity. 
Still assuming normal ordering, one gets : 
\ben
{\bf F_s} &=&
\sum_k \frac{i \; {\bf \nabla} \left[ {\bf d}.{\bf e}_k({\bf r}) \right] \, {\bf d}.{\bf e}_k({\bf r}) }
{\hbar} \int_0^{t-t_0} d\tau S^{(+)}(t) S^{(-)}(t-\tau) \; e^{-i(\omega_k -\omega_L)
\tau}  + h.c. \no \\
	&=& {\bf \nabla} \left[ \sum_k \frac{i \; ({\bf d}.{\bf e}_k({\bf r}) )^2}
{2 \hbar} \int_0^{t-t_0} d\tau S^{(+)}(t) S^{(-)}(t-\tau) \; e^{-i(\omega_k -\omega_L)
\tau}  \right]+ h.c.
\label{eq:Fs}
\een

In order to get the physical meaning of this integral, 
let us write the Heisenberg equation for 
$S^{(-)}$ and use the expression of $a_k$ in order to get :
\beq
\frac{dS^{(-)}}{dt} = i \delta _L S^{(-)}  - 2 i   S^{(z)}
\frac{{\bf d}.{\bf A}_o}{\hbar} +  2 \sum _k \frac{({\bf d}.{\bf e}_k)^2}{\hbar^2}
\int_0 ^{t-t_0} d\tau S^{(z)}(t) S^{(-)}(t-\tau) e^{-i(\omega_k -\omega _L)\tau} ,
\label{eqsm}
\eeq
where $\delta _L= \omega _L- \omega _o$.
The integral appearing here has the same structure as the one appearing 
in eq.~\ref{eq:Fs} under the gradient, and corresponds to the 
well known relaxation and light shift 
terms in the evolution of the dipole components. It can be calculated as usual 
using a Markov approximation \cite{CDG}, which is possible here because as said above 
we assumed that the cavity features are wide compared to the free-space 
relaxation rate of the atom. In the weak excitation limit, which is assumed here, one can 
approximate the evolution of the correlation functions by their free evolution 
during the short memory time of the reservoir. Note that an extension should 
be made to the strong excitation regime by considering relaxation in a dressed 
state basis \cite{DC}. Making the approximation $S^{(-)}(t-\tau) = S^{(-)}(t) \, e^{-i(\omega_L-\omega_0) \tau}$
during the correlation time \cite{CDG}, one can define the quantities : 
\ben
\sum_k \frac{({\bf d.e}_k({\bf r}) )^2}{\hbar^2 } \int_0 ^{t-t_0} d\tau  \; \;
e^{-i(\omega_k -\omega _o)\tau} &=& 
\sum_k \frac {({\bf d.e}_k({\bf r}) )^2}{\hbar^2 } 
\left( \pi \delta (\omega_k -\omega _o)
 + i {\cal P}\left( \frac{1}{\omega_k-\omega_o} \right) \right)  \no \\
&=& \Gamma ({\bf r})/2 + i \Delta  ({\bf r})
\een
where one has therefore :
\ben
\Gamma ({\bf r}) &=& \frac{2\pi}{\hbar^2 } \sum_k ({\bf d.e}_k({\bf r}) )^2 
\delta (\omega_k - \omega _o)   \label{tg} \\
\Delta  ({\bf r}) &=& \sum_k \frac{({\bf d.e}_k({\bf r}) )^2}{\hbar^2}
  {\cal P}\left( \frac{1}{\omega_k-\omega _o } \right) \label{td}
\een
Using these definitions and $S^{(z)} S^{(-)} = - S^{(-)}/2$, eq. \ref{eqsm} becomes as expected
\beq
\frac{dS^{(-)}}{dt} = (i \delta _L - i \Delta  ({\bf r}) - \frac{\Gamma ({\bf r})}{2}) \; S^{(-)}  - 2 i   S^{(z)}
\frac{{\bf d}.{\bf A}_o}{\hbar}
\eeq
so that the atomic frequency $\omega_0$ becomes $\omega_0 + \Delta  ({\bf r})$.
We note that the free-space value of $\Delta({\bf r})$
is a diverging quantity, which is 
usually assumed to be absorbed in the definition of the atomic levels; therefore, 
one considers here only the (finite)  change of $\Delta({\bf r})$ 
with respect to this free-space 
value, that will be denoted $\Delta'({\bf r})$ :
\beq
\Delta'({\bf r})= \Delta_{cav}({\bf r}) - \Delta_{vac} ({\bf r})
\eeq 
In the expression of the extra term in the force, 
one can use either $\Delta'({\bf r})$ or 
$\Delta_{cav}({\bf r})$, since the gradient of  $\Delta_{vac}({\bf r})$ 
is zero anyway, and one obtains finally~:
\ben 
{\bf F_s}&=& \hbar/2 \; \left( S^{(+)} S^{(-)} \;  {\bf \nabla} (-\Delta'({\bf r})  + 
i \Gamma ({\bf r})/2) +  {\bf \nabla} (-\Delta'({\bf r})
 -i \Gamma ({\bf r})/2) \; S^{(+)} S^{(-)} \right) \no \\
&=& -\, {\bf \nabla} (\hbar \, \Delta'({\bf r})) \; \; \Pi_e
\label{eqf}
\een
where $\Pi_e = S^{(+)} S^{(-)} = S^{(z)} + 1/2$ is the excited state population. 
Therefore the gradient of the cavity-induced light shift creates a force, 
which can attract the atom towards the cavity center for an appropriate choice of 
the atom-cavity detuning. 
Physically, it makes sense that ${\bf F_s}$ is also proportionnal 
to the excited level population. 
In order to characterize more precisely the behaviour of this force, 
one needs now the explicit space 
dependence of $ \Delta'({\bf r}) $. The same calculation will also give
the space-dependent damping $\Gamma ({\bf r})/2$, which will be
useful for calculating the steady state and evolution of the atomic operators.

\section{Cavity-induced  relaxation rates and light shifts}

For definiteness, we will consider the case of
 a spherical cavity of radius R and 
of reflectivity and transmittivity coefficients $\rho$ and $\tau$, 
with $\rho^2 + \tau^2 = 1$, and $\tau^2 = T$.
 We will  assume that $kR \gg 1$ (typically $kR =10^5$), and a moderate cavity 
finesse (in the  range 10-100). A crucial parameter is the solid angle 
subtended by the cavity, that will be denoted $\Delta \Omega_{cav}$. 
For instance, a cavity half-aperture angle of $45$ degrees 
 gives  ${\Delta \Omega_{cav} \over 4 \pi} = 0.3$, and therefore
${\Delta \Omega_{vac} \over 4 \pi} = 0.7$ as the fraction of space still
occupied by vacuum modes. All parameters quoted above seem accessible 
from an experimental point of view, and we will show now
that they allow one to get quite significant cavity-induced effects.

In order to calculate the explicit space and frequency dependence of 
the relaxation rate and level shift, we have followed two 
parallel approaches, which are described in detail in another 
publication \cite{JMD1}. The first one is to solve explicitly the field equations in a spherical
geometry, taking into account the considerable simplifications which appear since
we are interested in the field at distances smaller than $\sim 100\lab$ off the origin.
In geometric optics this involves light-rays with an impact parameter smaller 
than $100 \lab$, and therefore, in
the multipole expansion of the field, it will be sufficient to 
consider harmonics up to order $l\le 100$. 
Actually, up to $300$ harmonics have be used in order
to check consistency. Another very important point is that 
the continuity equations of the fields on the mirrors must be expressed
at a distance $R \simeq$ 1 cm. With $\la =780$ nm, this
corresponds to $kR \simeq 80000$, which is quite large, and allows one to 
use asymptotic forms of the solutions on the 
mirrors. Under these assumptions, it can be shown that 
the expression of $ \Delta'({\bf r}) $ and $\Gamma ({\bf r})/2$
can be given an explicit operatorial form in the space of mode
functions, and then evaluated numerically in a spherical harmonics basis \cite{JMD1}.

Besides this ``exact" calculation, we have also looked for an approximate
solution, inspired by ray-optics considerations, and eventually checked by comparison
with the complete numerical calculation. 
Using this cross-checking method, we obtain finally 
that the effect of the cavity can be described
to a very good approximation by the following formulas:
\beq
{\Gamma ({\bf r}) \over \Gamma_{vac}}=
\int_{4 \pi} {d\dir\over 4\pi} \; \;{3\over 2}  (1 - ({{\bf d}.\dir \over d})^2) 
\big( {T\over \be{1- \rho e^{2i \phi}}^2 }\cos^2\bigl(k\dir.\rr) \; +\;
{T\over \be{1+ \rho e^{2i \phi}}^2} \sin^2\bigl(k\dir.\rr) \big)
\label{fg}
\eeq
\beq
{\Delta' ({\bf r}) \over \Gamma_{vac}}=
\int_{4 \pi} {d\dir\over 4\pi}\; \;{3\over 2}  (1 - ({{\bf d}.\dir \over d})^2) 
\big( {\rho \; \sin(2 \phi)\over \be{1- \rho e^{2i \phi}}^2 }\cos^2\bigl(k\dir.\rr) \; - \;
{\rho \; \sin(2 \phi)\over \be{1+ \rho e^{2i \phi}}^2} \sin^2\bigl(k\dir.\rr) \big)
\label{fd}
\eeq
where the notation $\dir$ describes a direction in space, while $\phi$ is 
a cavity detuning parameter that will be detailed below. These expressions have
a straighforward interpretation, because they appear basically as integrals over the 
direction of light rays : in the integral over the directions,
$\rho$ is the mirror reflectivity for rays subtended by the cavity, and 
is zero for rays outside the cavity solid angle. The different factors 
appearing in the integrals can be interpreted in the following way~:

\begin{itemize}

\item The first factor under the integral
corresponds to polarisation effects, taking into account the transverse
character of the field. 

\item The second (resonance) factor is of the usual Fabry-Perot
form, where $\phi$ is the cavity phase shift which includes first
a term $\phi_0 = \omega_0 R/c$. The complete calculation shows that,
in order to obtain a correct result outside the cavity center, $\phi$ should
include also a contribution from spherical aberrations, that is :
$\phi = \phi_0 + \frac{k (r^2 - (\dir.\rr)^2)}{2 R}$. This second term 
corresponds to the extra phase shift experienced by rays
going through point $\rr$ while  propagating  along the $\dir$ direction.
The resonance factor has different expressions
for the damping and the lamb shift
terms, which correspond respectively to the active and reactive parts
of the coupling. This is clearly apparent from the integrals of eq. \ref{tg} and \ref{td}, 
which involve either a delta function or a principal part. 
In the first case, the integration is trivial, and yields
 the resonance term of eq. \ref{fg}, while in the second case
the result is obtained by contour integration \cite{JMD1}, and gives the (dispersive) 
second term of eq. \ref{fd}.

\item The third term under the integrals 
is the stationnary wave pattern corresponding either 
to odd modes (with an anti-node in the center and a $\cos^2\bigl(k\dir.\rr)$
space dependence) or to even modes (with a node in the center and a 
$\sin^2\bigl(k\dir.\rr)$ space dependence). 

\item Finally, the integration over the mirrors is conveniently performed in spherical 
coordinates, by taking the $z$ axis along the cavity axis, and varying the 
azimuthal angle $\theta$ from $0$ to $\theta_{mirror}=\theta_{m}$. 
Improved accuracy (better than $1 \%$) is obtained
if one takes into account the fact that the rays which would 
be reflected near the edge of the mirror are actually lost due to diffraction
and fail to do as many round-trips as the other ones. We have shown \cite{JMD1} that
this effect can be taken into account very simply 
by decreasing $\theta_{m}$ to
$\theta_{eff} = \theta_{m} - \delta \theta$, with 
$\delta \theta = 1 / \sqrt{k R T}$ for symmetrical mirrors. 

\end{itemize}

\begin{figure}[ht]
\includegraphics[width=8cm]{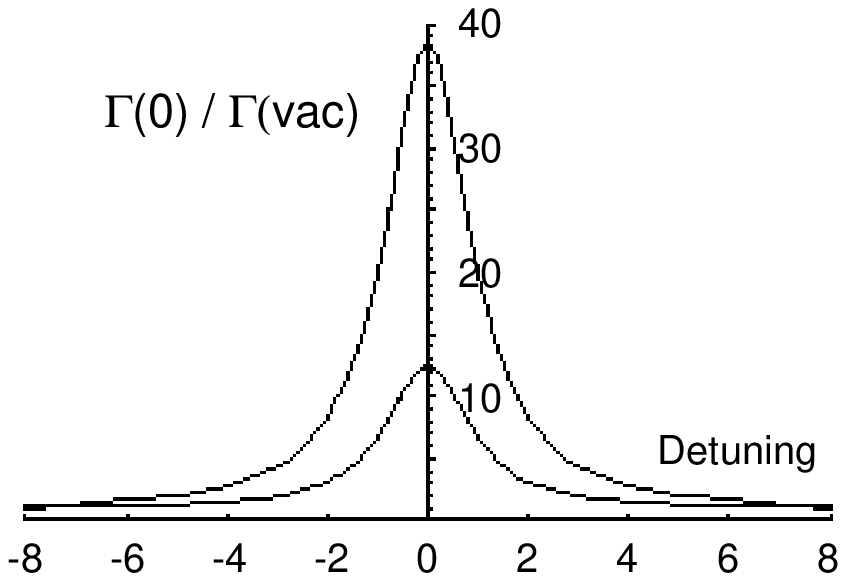} 
\includegraphics[width=8cm]{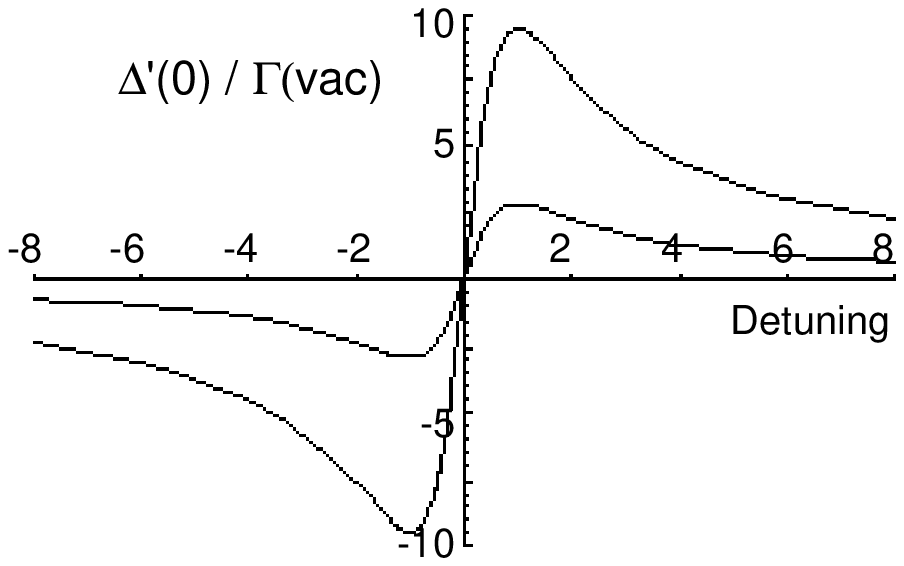} 
\caption{ 
Normalized damping $\Gamma(0) /\Gamma_{vac}$ (left) and level shift 
$\Delta'(0) /\Gamma_{vac}$ (right) at the cavity center, as a function of 
the atom-cavity detuning $(\omega_0-\omega_{cav})$ normalized to the cavity linewidth. 
The amplitude reflection coefficient of the mirrors is taken to be $\rho=0.98$,
and the numerical aperture of the cavity is 0.7.
The upper curves correspond to  a dipole oriented perpendicular to the cavity axis, 
and the lower curves to a dipole oriented along the cavity axis.}
 \label{ad}
\end{figure}

The first results which can be obtained from the previous formulas are obviously
the shift and damping at the cavity center, as a function of the atom-cavity
detuning. For a dipole orientation parallel to the cavity axis, we obtain 
straightforwardly :
\beq
{\Gamma_{par} ({\bf 0}) \over \Gamma_{vac}}= 
{\Delta \Omega_{vac} \over 4 \pi} (1 + {\sin^2\theta_m \over 2})+
{\Delta \Omega_{cav} \over 4 \pi} (1 - {\cos\theta_m (1+\cos\theta_m) \over 2})  
 {T\over \be{1 - \rho e^{2i \phi_0}}^2 }
\eeq
\beq
{\Delta'_{par} ({\bf 0}) \over \Gamma_{vac}}=
{\Delta \Omega_{cav} \over 4 \pi} (1 - {\cos\theta_m (1+\cos\theta_m) \over 2})  
{\rho \; \sin(2 \phi_0)\over \be{1 - \rho e^{2i \phi_0}}^2 }
\eeq
while for a dipole orientation perpendicular to the cavity axis, we have :
\beq
{\Gamma_{perp} ({\bf 0}) \over \Gamma_{vac}}= 
{\Delta \Omega_{vac} \over 4 \pi} (1 - {\sin^2\theta_m \over 4})+
{\Delta \Omega_{cav} \over 4 \pi} (1 + {\cos\theta_m (1+\cos\theta_m) \over 4})  
 {T\over \be{1 - \rho e^{2i \phi_0}}^2 }
\eeq
\beq
{\Delta'_{perp} ({\bf 0}) \over \Gamma_{vac}}=
{\Delta \Omega_{cav} \over 4 \pi} (1 + {\cos\theta_m (1+\cos\theta_m) \over 4})  
{\rho \; \sin(2 \phi_0)\over \be{1 - \rho e^{2i \phi_0}}^2 }
\eeq
We note that these expressions yield for a randomly oriented dipole :
\beq
{\Gamma_{av} ({\bf 0}) \over \Gamma_{vac}}= 
{\Delta \Omega_{vac} \over 4 \pi}+
{\Delta \Omega_{cav} \over 4 \pi} {T\over \be{1 - \rho e^{2i \phi_0}}^2 },\;\;\;
{\Delta'_{av} ({\bf 0}) \over \Gamma_{vac}}=
{\Delta \Omega_{cav} \over 4 \pi} 
{\rho \; \sin(2 \phi_0)\over \be{1 - \rho e^{2i \phi_0}}^2 }
\eeq
which have a straightforward interpretation in terms of resonant enhancement
of the rays subtended by the cavity.
We note that these results are the same as those given in ref.~\cite{HCTF}, up to factor two
resulting from the fact that this reference was considering spatially averaged
values rather than the peak value at the cavity center (see below for the space dependence).
These functions are plotted on fig. \ref{ad} for ${\Omega_{cav} \over 4 \pi} = 0.3$ and 
$\rho = 0.98$. It can be seen that very significant effects
occur for these quite reasonable parameters, yielding more than 30-fold increase 
in the damping rate at the cavity center. 

\begin{figure}[ht]
\includegraphics[width=8cm]{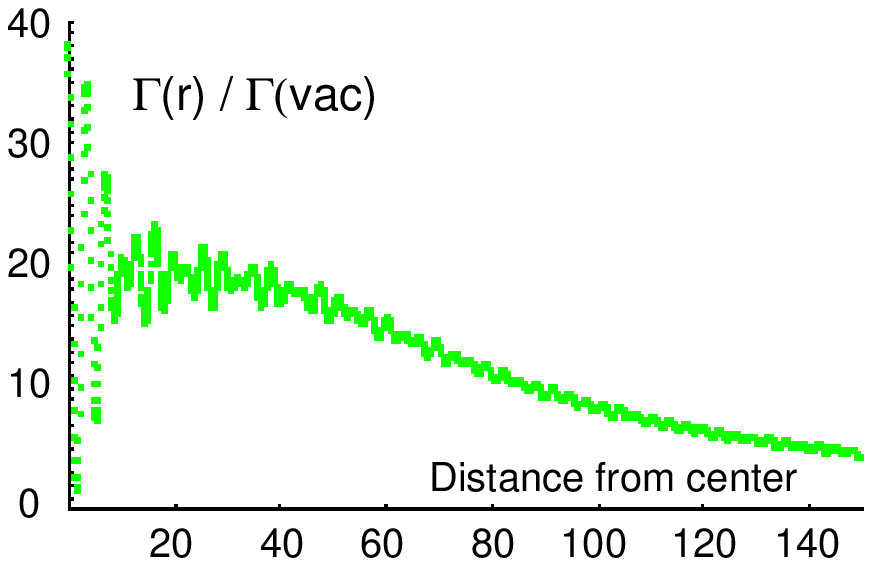} 
\includegraphics[width=8cm]{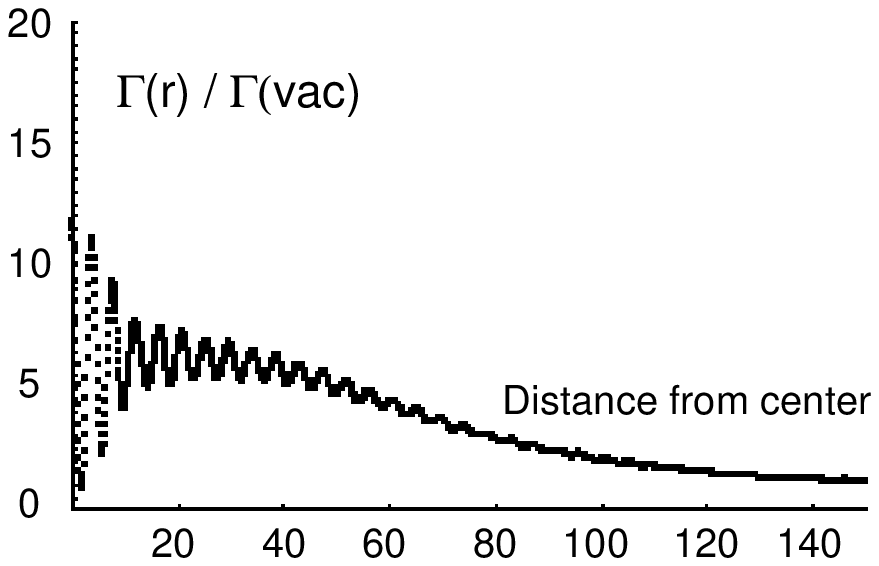} 
\includegraphics[width=8cm]{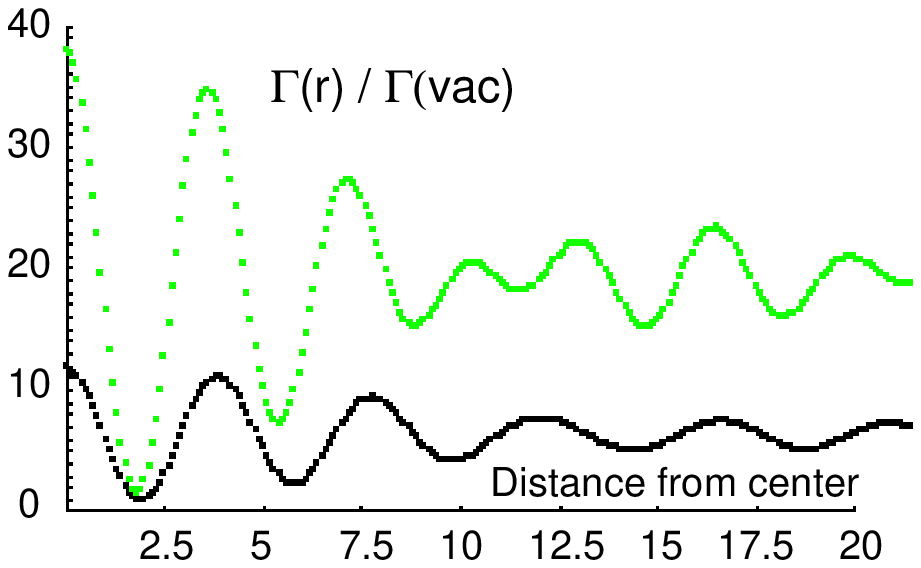} 
\caption{ 
Spatial variation of the atomic damping $\Gamma(r) /\Gamma_{vac}$ on the cavity axis, obtained
from the analytical formula given in the text. The atom-cavity detuning is taken equal to
zero at the cavity center (see Fig. \ref{ad}), and the horizontal axis unit is $1/k$.
The amplitude reflection coefficient of the mirrors is $\rho=0.98$,
and the numerical aperture of the cavity is 0.7.
The upper left curve corresponds to  a dipole oriented perpendicular to the cavity axis, 
the upper right curve to a dipole oriented along the cavity axis,
and the lower curve is a zoom close to the cavity center.}
 \label{sp}
\end{figure}

We can then look at the results as a function of space for a given frequency,
which are essential for the present paper. Two 
atom-cavity detunings are specially worth looking at : 
the resonant frequency at the cavity center,
which yields maximum change in the damping rate but no cavity shift, and  frequencies
detuned by plus or minus
 half a cavity linewidth, which yield maximum cavity shifts.
As an example, the results for the damping rates are  plotted on fig. \ref{sp}.
The general values of the damping and level shifts for arbitrary values of the reflectivities
$\rho_1$, $\rho_2$ of the two mirrors are given in \cite{JMD1}, Appendix C. 

\section{Discussion}

From the results of the previous section one can deduce
some features of the ``vacuum-induced" force given by eq.~\ref{eqf}.
For simplicity, we consider an atom at point $\bf{r}$
with zero velocity. The simplest configuration in which this
force should be dominant is a weak stationnary wave resonant
on the atom, but detuned from the cavity. In that case,
both the usual scattering and dipole force are zero, while the cavity-induced 
light shift is maximum for a detuning of about half a cavity width.
It can be seen easily that 
the force is attractive when the cavity is detuned to 
the blue side of the atomic resonance $(\omega_0 < \omega_{cav})$ . This can be understood
physically by realizing that the cavity ``repels" the excited state level,
and that this level has to go down in order to get an attractive potential.
The corresponding potential wells are a few $\hbar \Gamma$ (see eq.~\ref{eqf}),
which is quite significant if cold atoms are used.

Besides the effect described here in the weak 
excitation regime, another possibility is to couple a laser beam 
inside the cavity, with a red atom-laser detuning, 
in order to create a dipole trap attracting the 
atom towards the cavity center.  
A very interesting effect can then be expected : 
since the atomic relaxation will be modified by the cavity, 
the balance between 
the trapping and heating effects of the dipole trap will be changed 
with respect 
to free space, enabling new regimes where the dipole force can be both 
attractive and damped.
In that case, the calculation above should be extended 
to moving atoms (non-zero velocity) in order to calculate 
friction forces, as well as to calculate the diffusion coefficient. 
For very cold atoms, the quantum aspect of motion should also be included;
this could be done in principle since
we have obtained the explicit space dependence of the trapping potential.

\section{Conclusion}

As a conclusion, we have shown that macroscopic cavities with large 
numerical apertures are interesting candidates for cavity QED experiments in 
the optical domain. The possibility to use light-induced force to hold the atom 
close to the cavity center is quite attractive, and it offers the possibility to study 
a well defined quantum system, including its external degrees of freedom.
The expressions that we have obtained are general, and by using the results
of ref.~\cite{JMD1} they can be applied 
to any kind of low-finesse cavity, including the case of a ``half-cavity"
with only one mirror imaging the dipole onto itself \cite{inn1,inn2,inn3}.

\end{document}